\documentclass[conference]{IEEEtran}

\usepackage{cite}
\usepackage{siunitx}
\usepackage{graphicx}
\usepackage[tight,footnotesize]{subfigure}
\usepackage{pifont}
\usepackage{amssymb}
\usepackage{dblfloatfix}
\usepackage{booktabs}
\usepackage{amsmath}
\usepackage{upgreek}

\begin{document}
\bstctlcite{IEEEexample:BSTcontrol}

\title{Design-oriented Modeling of 28\,nm FDSOI CMOS Technology down to 4.2\,K for Quantum\,Computing\vspace{-0.7cm}}

\author{\IEEEauthorblockN{Arnout Beckers\IEEEauthorrefmark{2},
Farzan Jazaeri\IEEEauthorrefmark{2},
Heorhii Bohuslavskyi\IEEEauthorrefmark{3}, 
Louis Hutin\IEEEauthorrefmark{3}, Silvano De Franceschi\IEEEauthorrefmark{3}, and
Christian Enz\IEEEauthorrefmark{2}}
\IEEEauthorblockA{\IEEEauthorrefmark{2}Integrated Circuits Laboratory (ICLAB), Ecole Polytechnique F\'ed\'erale de Lausanne (EPFL), Switzerland,\\
\IEEEauthorrefmark{3}CEA-L\'eti, Grenoble, France\\}
arnout.beckers@epfl.ch
\vspace{0.4cm}}

\maketitle

\begin{abstract}
In this paper a commercial 28-nm FDSOI CMOS technology is characterized and modeled from room temperature down to 4.2\,K. Here we explain the influence of incomplete ionization and interface traps on this technology starting from the fundamental device physics. We then illustrate how these phenomena can be accounted for in circuit device-models. We find that the design-oriented simplified EKV model can accurately predict the impact of the temperature reduction on the transfer characteristics, back-gate sensitivity, and transconductance efficiency. The presented results aim at extending industry-standard compact models to cryogenic temperatures for the design of cryo-CMOS circuits implemented in a 28\,nm FDSOI technology.
\end{abstract}
\IEEEpeerreviewmaketitle
\let\thefootnote\relax\footnote{\hspace*{-1em}This project has received funding from the European Union's Horizon 2020 Research \& Innovation Programme under grant agreement No. 688539 {MOS-Quito}.}
\section{Introduction}
Quantum computing can reshape many fields by providing the computing power to solve exponentially-growing problems. To reach this computing power, an important challenge today is the scale-up to larger qubit numbers\cite{almudever_engineering_2017,vandersypen_interfacing_2017}. Each additional qubit adds to the complexity of the room-temperature control-equipment, introducing more interconnections, wiring capacitance and thermal-noise pickup. Implementing the control equipment in miniaturized cryo-CMOS represents an interesting solution for scalability and fast qubit-information processing. Digital, analog and RF circuits are then required to operate down to deep-cryogenic temperatures. In this context, silicon-on-insulator technology provides an excellent platform to create a complete, scalable quantum computing system\cite{ekanayake_characterization_2010,franceschi_soi_2016}. CMOS-compatible spin qubits have been developed in a FDSOI nanowire technology \cite{maurand_cmos_2016,hutin2016si}, and can be co-integrated with the required qubit-control circuits designed in a FDSOI CMOS process. Increased digital/analog performances down to 4.2\,K have been reported for advanced FDSOI CMOS technologies\cite{bohus,balestra2017physics}. The suitability of the 28\,nm node for qubit-control electronics has been reported in \cite{bohus}. Furthermore, the FDSOI back-gate offers a versatile tool to control the power consumption close to the qubits \cite{bohus}, which may become essential to keep decoherence at bay in co-integrated quantum-classical circuits. Due to a strong reduction of the cooling power at millikelvin temperatures, the control circuits are envisioned to operate at 4.2\,K or higher\textemdash and eventually the qubits as well\cite{vandersypen_interfacing_2017}. However, designing the cryo-CMOS qubit-control circuits\textemdash while meeting a stringent speed-power trade-off\textemdash is not an easy task, since cryo-CMOS device models are currently lacking in circuit-simulators. In order to keep pace with the rapid qubit-developments, device models remaining physically accurate down to 4.2\,K  are urgently required. To this purpose, important physical phenomena at deep-cryogenic temperatures,\,i.e. interface trapping and incomplete ionization, need to be included. Here, after a brief discussion of the measurement results down to 4.2\,K, the temperature-trends of the main technological parameters are investigated (Sec.\,\ref{sec:charac}). To explain some of these trends, we examine the temperature dependencies of interface trapping and incomplete ionization in the device physics of this technology (Sec.\,\ref{sec:physics}). It will be evidenced that these phenomena can respectively alter the slope factor and the threshold voltage. As we will demonstrate, adjusting the corresponding model parameters in the design-oriented simplified EKV model, allows to model the DC cryogenic device-performance down to 4.2\,K, including the effect of the back-gate (Sec.\,\ref{design-oriented}). These results pave the way towards the extension of industry-standard compact models down to the deep-cryogenic temperature regime \cite{poiroux_leti-utsoi2.1:_2015,bsimimg}. This will allow to explore optimal cryo-CMOS circuit designs for multiple applications, in particular qubit-control systems. 
\begin{table}[t]
	\centering
	\caption{Investigated Devices (28-nm FDSOI CMOS Process)}
	\label{table}
	\resizebox{0.24\textwidth}{!}{%
		\begin{tabular}{rcc}
			\toprule
			Symbol & Type & W/L\\ 
			\midrule
			{\tiny\ding{108}} & $n$MOS & 1\,$\upmu$m / 1\,$\upmu$m\\
			{\tiny\ding{115}} & $p$MOS & 1\,$\upmu$m / 1\,$\upmu$m\\	
			{\tiny\ding{116}} & $n$MOS & 1\,$\upmu$m / 28\,nm\\
			{\tiny\ding{110}} & $n$MOS & 210\,nm / 28\,nm\\
			{\tiny\ding{117}} & $n$MOS & 80\,nm / 28\,nm\\ 
			\bottomrule
	\end{tabular}}
\end{table}
\begin{figure*}[t]
	\centering
	\includegraphics[width=\textwidth]{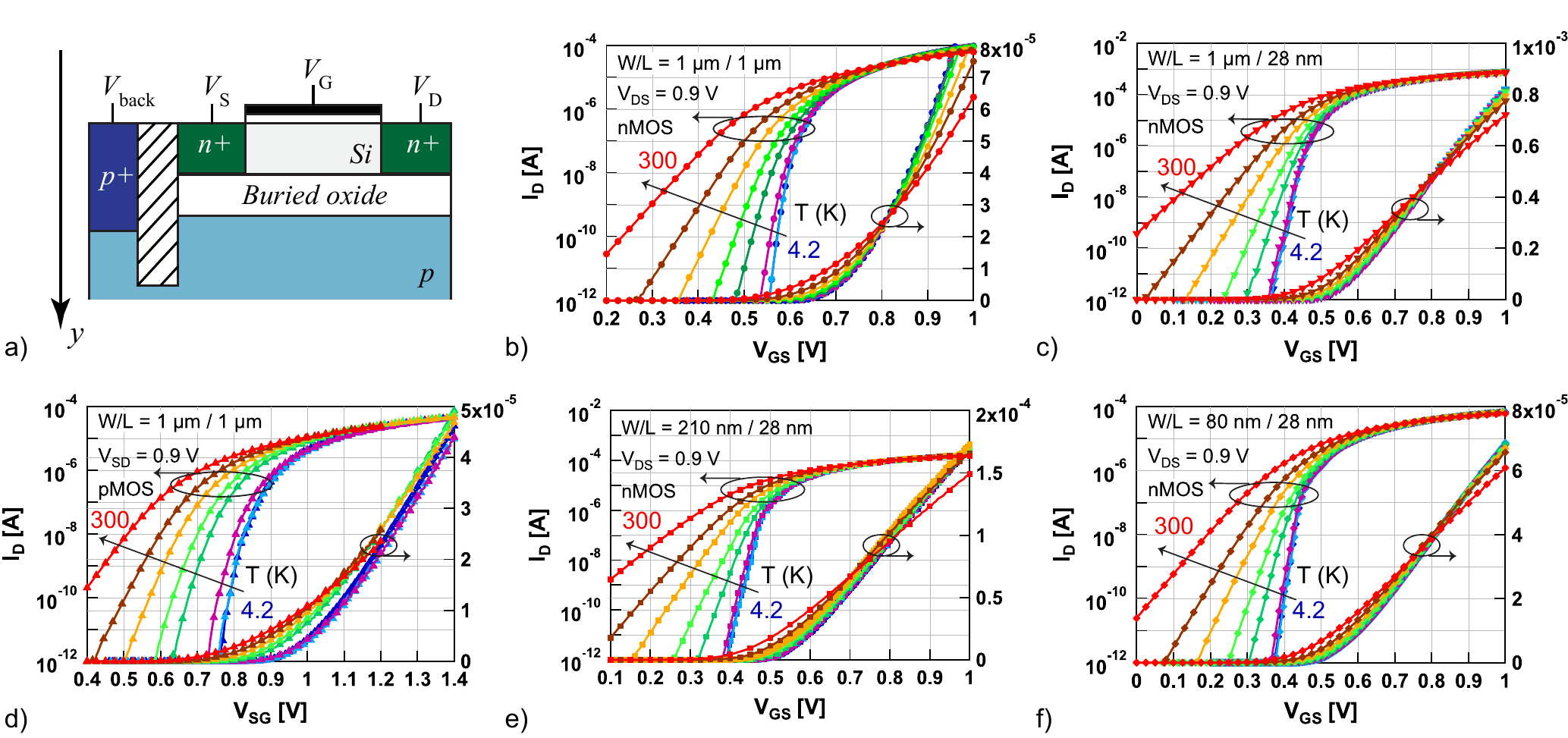}
	\vspace{-0.6cm}
	\hfil
	\caption{a) Schematic cross-section of a $n$MOS device fabricated in a 28\,nm FDSOI CMOS process\cite{bohus}, b)-f) Transfer characteristics measured in saturation  ($\vert V_{DS}\vert $ = 0.9\,V) with $V_{back}$=\,0\,V, at 300, 210, 160, 110,  77. 36, 10, and 4.2\,K for b) $n$MOS $W/L$\,=\,1 $\upmu$m\,/\,1\,$\upmu$m, c) $n$MOS $W/L$\,=\,1 $\upmu$m\,/\,28\,nm, d) $p$MOS $W/L$\,=\,1 $\upmu$m\,/\,1\,$\upmu$m, e) $n$MOS $W/L$\,=\,210\,nm\,/\,28\,nm, and f) $n$MOS $W/L$\,=\,80\,nm\,/\,28\,nm. The EOT for $n$MOS is 1.55 nm, and for $p$MOS 1.7 nm. The gate-source voltage is increased with a step size of $\vert V_{GS}\vert =$\,5\,mV. Colors indicate the temperature, and markers the device according to Table\,\ref{table}.}
	\label{fig:meas}
\end{figure*}
\section{\label{sec:charac}Measurements and Characterization}
\begin{figure*}[t]
	\includegraphics[width=\textwidth]{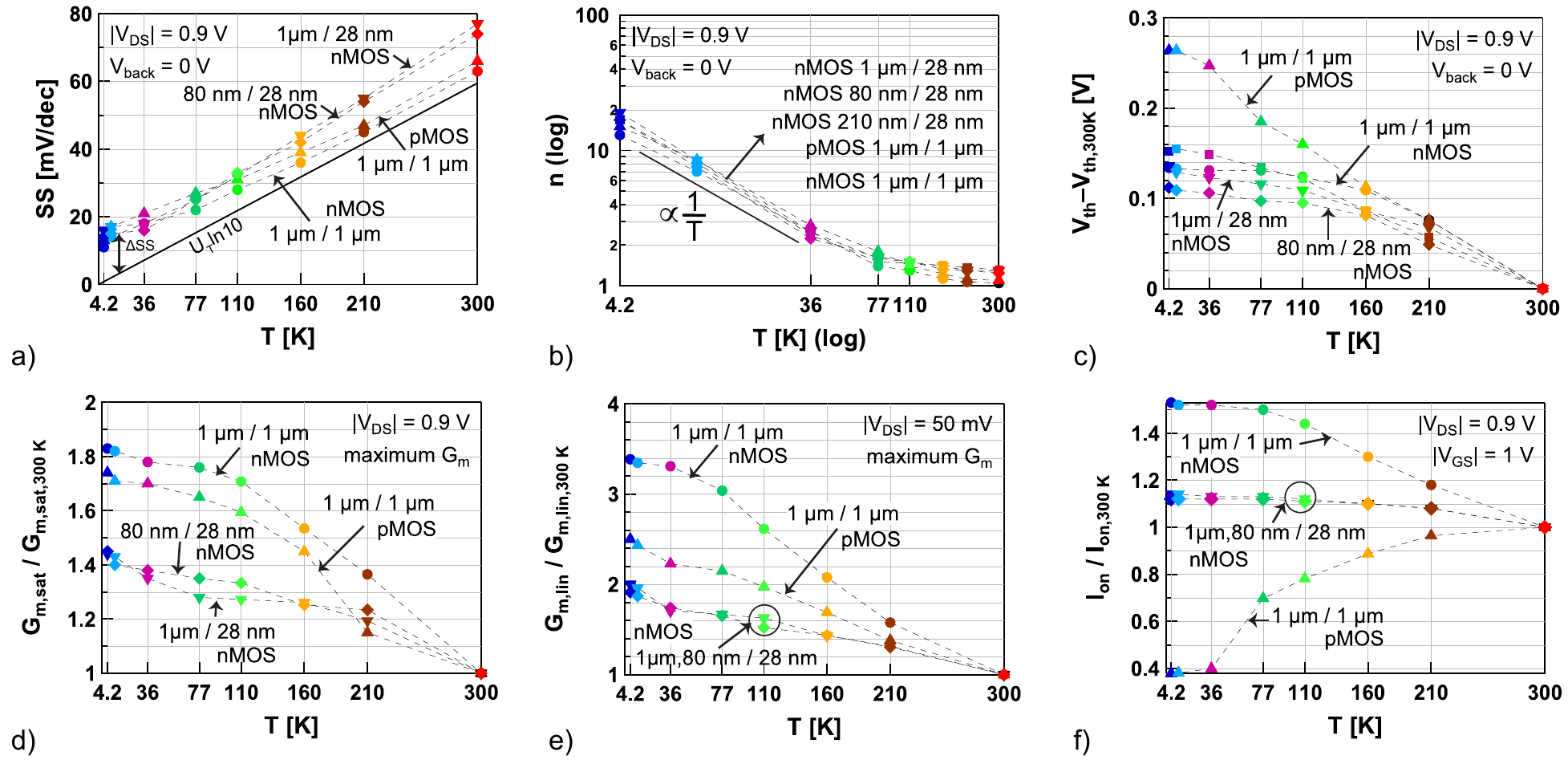}
	\vspace{-0.6cm}
	\caption{Technological parameter extraction at 300, 210, 160, 110,  77. 36, 10, and 4.2\,K, a) Subthreshold swing,  modeled for long-channel devices by $SS=n_0U_T\ln10+\Delta SS$, with $\Delta SS \propto N_{it}$ \cite{beckers_jeds_2018,tedpaper}, b) Slope factor in log-log scale highlighting its hyperbolic temperature dependency, c) Threshold-voltage shift with respect to 300\,K, d)-e) Maximum transconductances in saturation and linear regions of operation, normalized to 300\,K, f) On-state current normalized to 300\,K. Colors indicate the temperature, and markers the device according to Table\,\ref{table}.}
	\label{fig:char}
\end{figure*}

The set of devices under investigation, fabricated in a commercial 28\,nm FDSOI CMOS process, is listed in Table\,\ref{table}. Details on the fabrication procedure can be found in\cite{bohus}. A schematic cross-section of the FDSOI $n$MOS transistor is drawn in Fig.\,\ref{fig:meas}a. The transfer characteristics were measured in the linear ($V_{\mathrm{DS}}$\,=\,50\,mV) and saturation operational regimes ($V_{\mathrm{DS}}$\,=\,0.9\,V) at different temperatures\cite{bohus}. The back-gate voltage ($V_{back}$) is swept from $-$0.9\,V to 0.9\,V. Figs.\,\ref{fig:meas}b-f show the transfer characteristics in saturation at $V_{back}$=\,0\,V. The strong improvement of the $SS$ with decreasing temperature is clear for all devices in Fig.\,\ref{fig:meas}. However, note that the improvement from 36\,K down to 4.2\,K is marginal, especially for the short devices. Based on the measurements in Fig.\,\ref{fig:meas}, the following technological parameters have been extracted (Fig.\,\ref{fig:char}): the subthreshold swing ($SS$), slope factor ($n$), threshold voltage ($V_{th}$), transconductance in linear and saturation ($G_{m,lin},G_{m,sat} $), and the on-state current ($I_{on}$). As illustrated in Fig.\,\ref{fig:char}a, for temperatures below $\approx$\,160\,K, the extracted average $SS$-values show an increasing offset, $\Delta SS$, from the thermal limit, $U_T\ln10$, with $U_T\triangleq kT/q$ the thermal voltage. $\Delta SS$ reaches around 10\,mV/dec at 4.2\,K for long $n$MOS, since $U_T\ln10$ predicts $\approx$\,0.8\,mV/dec. The necessary slope factors to reach such high $SS$ are extracted in Fig.\,\ref{fig:char}b using $n=SS/(U_T\ln10)$. From this figure, a hyperbolic temperature-dependency of $n$ is evident, which is not strongly dependent on geometry at cryogenic temperatures. The data points below 77\,K in Fig.\,\ref{fig:char}a cannot be explained by $n_0U_T\ln10$ with a slope-factor $n_0$ limited by 2, according to $n_0=1+C_{dep}/C_{ox}$, where $C_{dep}$ is the depletion capacitance and $C_{ox}$ the oxide capacitance. Moreover, including the interface-trap capacitance, $C_{it}=qN_{it}$, i.e. $n_0=1+(C_{dep}+C_{it})/C_{ox}$ with $N_{it}$ the density-of-interface-traps per unit area, would lead to very high extracted values for $N_{it}$ in the order of $\SI{e13}{\per\square\centi\meter}$ at 4.2\,K\cite{elewa,trevisoli_junctionless_2016,hafez_assessment_1990}. However, it should be emphasized that in this $n_0$-formula the temperature-dependent occupation of interface-traps is not taken into account. This will be further investigated in Sec.\,\ref{sec:physics}. The shift in threshold voltage at 4.2\,K with respect to room temperature increases in the order of 0.1$-$0.3\,V (Fig.\,\ref{fig:char}c). Note that the largest $V_{th}$-increase is observed for $p$MOS, similarly to a 28-nm bulk process\cite{beckers1}. Furthermore, the maximum $G_{m,sat}$ and $G_{m,lin}$ (Figs.\ref{fig:char}d-e) improve down to 4.2\,K, e.g.\,respectively $\times$\,3.4 (linear) and $\times$\,1.8 (saturation) for $n$MOS $W/L$\,=\,1 $\upmu$m\,/\,1\,$\upmu$m. In Fig.\ref{fig:char}f, $I_{on}$ is extracted at $\vert V_{GS}\vert=$\,1\,V. Note that the actual trend of $I_{on}$ with temperature is strongly dependent on the bias and the device-type. At a standard supply-voltage of 1\,V, the on-state current increases with decreasing temperature for long $n$MOS (Fig.\,\ref{fig:meas}b), while it decreases for $p$MOS (Fig.\,\ref{fig:meas}d). The drain-induced barrier-lowering (DIBL) is approximately zero for the long devices. For the short devices, small improvements have been extracted down to 4.2\,K, but the short-channel effect remains largely temperature-independent,\,e.g. decreasing from 0.07\,V/V (300\,K) to 0.068\,V/V (4.2 K) for $n$MOS $W/L=$80\,nm\,/\,28\,nm, and from 0.065\,V/V (300\,K) to 0.059\,V/V (4.2\,K) for $n$MOS $W/L=$210\,nm\,/\,28\,nm (extracted at $I_D=$ \SI{e-8}{\ampere}). 
\begin{figure*}[t]
	\centering
	\includegraphics[width=\textwidth]{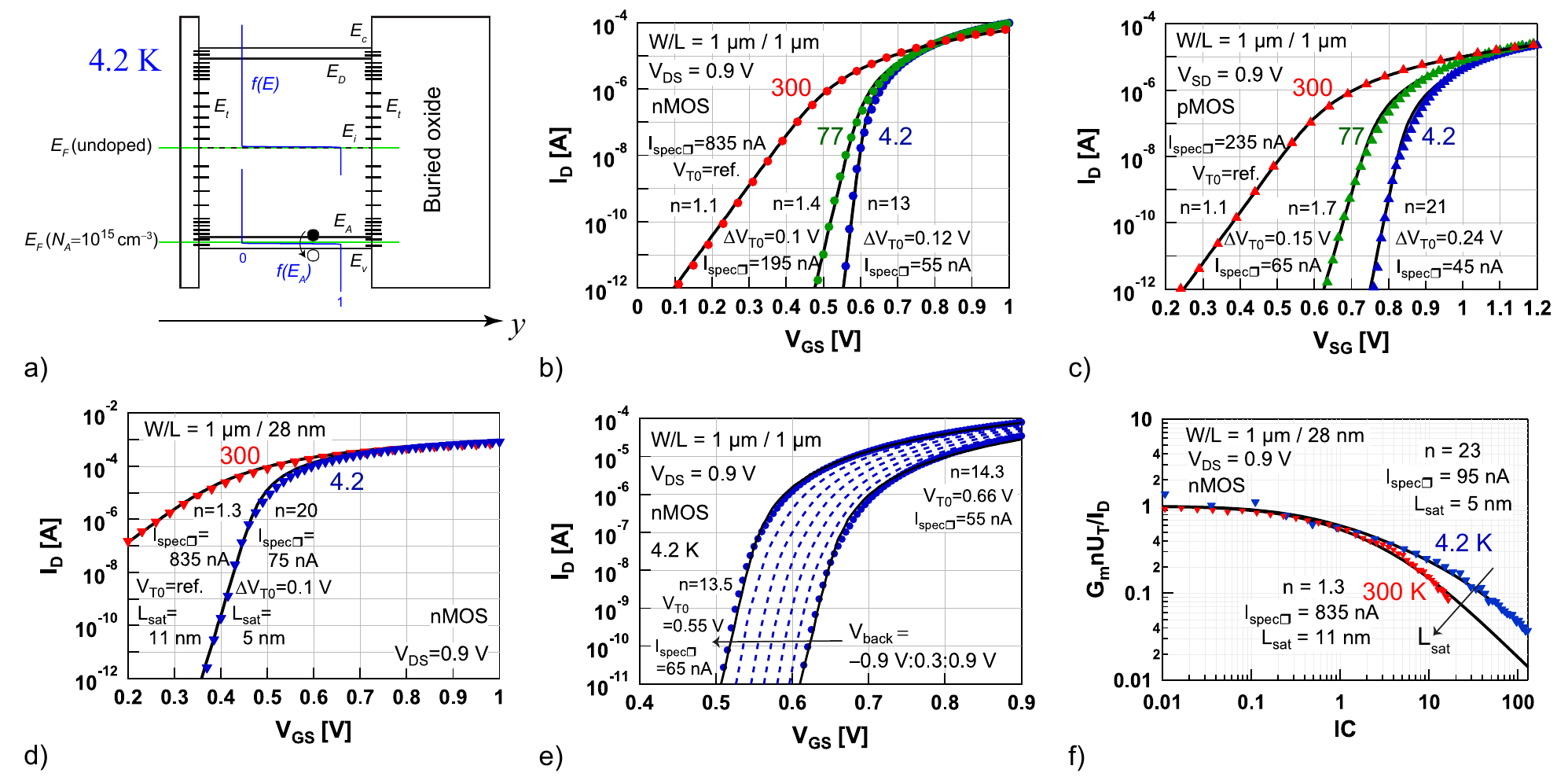}
	\vspace{-0.7cm}
	\caption{Band diagram along the $y$-direction in Fig.\,\ref{fig:meas}a. The position of the Fermi-level, $E_F$, is calculated at 4.2\,K in the case of undoped and low-doped channel, including incomplete ionization and bandgap widening at 4.2\,K. The simulated Fermi-Dirac distribution at 4.2\,K is plotted for both cases. The temperature-dependent occupation of the interface traps, $f(E_t)$, can degrade the subthreshold swing. Due to band bending of $E_A$ under $E_F$ at the front-gate, frozen-out impurities under the surface become completely ionized ($f(E_A)=1$), long before subthreshold is reached ($E_F\approx E_c-3U_T$) due to field-assisted ionization. b)-e) Simplified EKV (solid lines) for  b) $n$MOS $W/L$\,=\,1 $\upmu$m\,/\,1\,$\upmu$m, c) $p$MOS $W/L$\,=\,1 $\upmu$m\,/\,1\,$\upmu$m, and d) $n$MOS $W/L$\,=\,1 $\upmu$m\,/\,28\,nm at 300, 77, and 4.2\,K (markers) for $V_{back}$=0\,V. The model parameters, $n$, $V_{T0}$ (threshold voltage), $I_{spec\scriptsize{\square}}$ (specific-current-per-square), and $L_{sat}$ (saturation length), are shown. e) Back-gate sensitivity at 4.2\,K modeled with simplified EKV. The model is shown for $V_{back}$=$-$0.9\,V and 0.9\,V with solid lines. Markers and dashed lines indicate measurements at 4.2\,K, f) Normalized transconductance efficiency versus the inversion coefficient, $IC$, at 300 and 4.2\,K.}\label{fig:modeling}
\end{figure*}
\section{\label{sec:physics}Device Physics}
The improvements of ($SS$, $G_{m,sat}$, $G_{m,lin}$) and the increase in $V_{th}$ follow directly from the temperature-scaling of the Fermi-Dirac distribution function, $f(E)$. As illustrated in Fig.\,\ref{fig:modeling}a, at 4.2\,K $f(E)$ is almost a step function. Therefore, the conduction band needs to be bent further downward to create sufficient overlap with the conduction-band density-of-states to reach inversion, increasing $V_{th}$. Note that the subthreshold region happens only when $E_F$ lies within $\approx 3U_T$ of $E_c$. However, as indicated by $\Delta SS$ in the previous section, the turn-on rate predicted in this way is too steep. In what follows, we investigate the impact of incomplete ionization/freeze-out and interface trapping to explain this observation. Both phenomena are modeled by the Fermi-Dirac distribution as well, i.e. respectively $f(E_A)$ (for $p$-type doping) and $f(E_t)$, where $E_A$ and $E_t$ are the acceptor and trap energy levels. This is qualitatively shown in Fig.\ref{fig:modeling}a. In case the Si-channel (Fig.\,\ref{fig:meas}a) is truly undoped, incomplete ionization should not be considered. However, if there is a certain background doping, e.g. $p$-type $N_A=\SI{e15}{\per\centi\meter\cubed}$, these impurities can be frozen out in the flatband condition at 4.2\,K. As shown in Fig.\ref{fig:modeling}a, the calculated $E_F$-position for low-doped channel at 4.2\,K lies under $E_A$, leading to freeze-out ($f(E_A)\ll1$). Nonetheless, these impurities will become quickly ionized due to field-assisted ionization\cite{foty_impurity_1990}, when $E_A$ bends under $E_F$ near the surface of the front-gate. In the subthreshold region, when $E_F\approx E_c-3U_T$, complete ionization can be assumed. Therefore, incomplete ionization cannot lead to the subthreshold-swing degradation observed in the previous section. Note, however, that for non-zero doping concentrations incomplete ionization can yield a small change in the threshold voltage, due to a modification of the charge-neutrality and the resulting $E_F$-position\cite{tedpaper}. Fig.\,\ref{fig:modeling}a highlights that the temperature-dependency of interface-trap occupation, $f(E_t)$, may influence the turn-on rate of the device down to 4.2\,K. Similarly to the derivation for bulk MOSFET presented in \cite{tedpaper}, by including $f(E_t)$ the temperature dependency of $n \propto 1/U_T$ can be derived for the front-gate in FDSOI as well. This gives $SS=n(T)U_T\ln 10=n_0U_T\ln10+\Delta SS$, where $n_0$ is the slope factor without interface traps, and $\Delta SS$ the subthreshold-swing offset as observed in the previous section. $\Delta SS$ is given by $(qN_{it}/C_{ox})\ln10\left[g_t/(1+g_t)^2\right]$ with $N_{it}$ the density-of-interface-traps and $g_t$ the trap degeneracy factor. Note that in this model, $N_{it}$ does not become multiplied with $U_T$, resulting in reasonable extracted values for $N_{it}$ at 4.2\,K ($\approx 10^{11}$-$\SI{e12}{\per\centi\meter\squared}$) lower than \cite{hafez_assessment_1990,trevisoli_junctionless_2016}. The $\Delta SS$-offset starts to increase below $\approx$\,160\,K since the subthreshold region happens when $E_F$ lies closer to $E_c$, where $N_{it}$ is observed to be higher already at 300\,K (Fig.\,\ref{fig:modeling}a). From this section, we conclude that interface trapping strongly degrades the $SS$ through the hyperbolic temperature-dependency of $n$, and incomplete ionization slightly alters the $V_{th}$-increase for non-zero doping concentrations. These two parameters, $n$ and $V_{th}$, can be modified accordingly in design-oriented models to predict deep-cryogenic operation. In the next section we demonstrate this relying on the simplified EKV model.
\section{\label{design-oriented}Design-oriented Modeling}
The design-oriented simplified EKV model is described in detail in\cite{enz2017nanoscale}. The suitability of this model for FDSOI processes has been assessed at room temperature\cite{bib:pezzotta:eurosoi:2018}.  
Using this model, the effects of the temperature reduction down to 4.2\,K on the transfer characteristics, the back-gate sensitivity, and the transconductance efficiency are modeled. The model accurately predicts the transfer characteristics down to 4.2\,K for long (Figs.\,\ref{fig:modeling}b-c) and short devices (Fig.\ref{fig:modeling}d). The strong increase in the $n$ model parameter at 4.2\,K accounts for the interface-trapping phenomenon. The $V_{T0}$ model parameter captures the change in the threshold voltage due to Fermi-Dirac scaling and incomplete ionization. Note that the used values for $n$ and $V_{T0}$ correspond to the extracted values in Fig.\,\ref{fig:char}b and \ref{fig:char}c.  Furthermore, as illustrated in Fig.\,\ref{fig:modeling}e, changing the $V_{T0}$ model parameter allows to capture the effect of the back-gate also at 4.2\,K\cite{bib:pezzotta:eurosoi:2018}. A small adjustment of the slope factor is necessary as well, to account for the change in $SS$ induced by the back-gate. Fig.\ref{fig:modeling}f verifies that the $G_m/I_D$ design-methodology remains valid for a 28\,nm FDSOI technology down to 4.2\,K. The normalized transconductance efficiency, $G_mnU_T/I_D$, is plotted versus the inversion coefficient $IC \triangleq I_{D,sat}/I_{spec}$, with $I_{spec}=I_{spec\scriptsize{\square}}W/L=2(W/L)n\mu C_{ox}U_T^2$. The specific-current-per-square,  $I_{spec\scriptsize{\square}}$, decreases over one order of magnitude from room temperature down to 4.2\,K. For the short-channel device, the additional $L_{sat}$-parameter, which denotes the length of the channel in full velocity saturation, decreases down to 4.2\,K due to a reduced phonon scattering. 
\section{Conclusion}
In this work, the DC operation of a 28\,nm FDSOI technology is modeled down to 4.2\,K by means of the simplified EKV model. A study of the device physics is first performed including the temperature dependencies of interface trapping and incomplete ionization. We find that the Fermi-Dirac temperature dependency of interface-trap occupation explains the large degradation of the subthreshold swing at 4.2\,K. In case impurities are present, freeze-out is of minor importance for device operation thanks to field-assisted ionization. These results bring us one step closer to the realization of large-scale silicon-based quantum computing systems. 
\bibliographystyle{IEEEtran}
\bibliography{eurosoi}
\end{document}